\documentclass{elsart}
\usepackage{graphics}
\usepackage{graphicx}
\usepackage{epsfig}
\usepackage{amssymb}

\begin{document}
\begin{frontmatter}
\title{Social dilemmas in an online social network: the structure and evolution of cooperation}
\author[ad1,ad2]{Feng Fu},
\ead{fufeng@pku.edu.cn}
\author[ad1,ad2]{Xiaojie Chen},
\author[ad1,ad2]{Lianghuan Liu},
\author[ad1,ad2]{Long Wang\corauthref{cor1}}
\ead{longwang@pku.edu.cn}

\corauth[cor1]{Corresponding author. Fax: +86-010-62754388.}
\address[ad1]{Center for Systems and Control, College of Engineering,
Peking University, Beijing 100871, China}
\address[ad2]{Department of Industrial Engineering and Management, College of Engineering, Peking University, Beijing 100871, China}

\begin{abstract}
We investigate two paradigms for studying the evolution of
cooperation---Prisoner's Dilemma and Snowdrift game in an online
friendship network obtained from a social networking site. We
demonstrate that such social network has small-world property and
degree distribution has a power-law tail. Besides, it has
hierarchical organizations and exhibits disassortative mixing
pattern. We study the evolutionary version of the two types of
games on it. It is found that enhancement and sustainment of
cooperative behaviors are attributable to the underlying network
topological organization. It is also shown that cooperators can
survive when confronted with the invasion of defectors throughout
the entire ranges of parameters of both games. The evolution of
cooperation on empirical networks is influenced by various network
effects in a combined manner, compared with that on model
networks. Our results can help understand the cooperative
behaviors in human groups and society.
\end{abstract}

\begin{keyword}
Social networks \sep Prisoner's Dilemma \sep Snowdrift game \sep
Cooperation \sep Network effects

\PACS 89.75.Hc \sep 87.23.Ge \sep 02.50.Le.
\end{keyword}
\end{frontmatter}

\section{Introduction}
Cooperative behaviors (mutualism and altruism) are ubiquitous in
human society as well as in virtual online community. For
instance, people unselfishly and collaboratively recommend
commodities such as books, songs, CD/DVDs, etc to each other.
Accordingly, this cooperative behavior (collaborative
recommendation) promotes the long tail which is the success
foundation of Amazon and eBay~\cite{Anderson}. And yet, according
to Darwinism, natural selection is based on competition. How can
natural selection lead to cooperation among selfish individuals?
Fortunately, together with classic game theory, evolutionary game
theory provides a systematic framework for investigating the
emergence and maintenance of cooperative behavior among unrelated
and selfish individuals. Two simple games, namely, Prisoner's
Dilemma game (PDG) and Snowdrift game (SG), as metaphors for
studying the evolution of cooperation have been extensively
adopted by researchers from different
background~\cite{Maynard82,Nowak92,Turner99,Doebeli05,Santos05,Szabo_review}.
In the original PDG, two players simultaneously decide whether to
cooperate (C) or to defect (D). They both receive $R$ upon mutual
cooperation and $P$ upon mutual defection. A defector exploiting a
C player gets $T$, and the exploited cooperator receives $S$, such
that $T>R>P>S$. As a result, it is best to defect regardless of
the co-player's decision. Thus, in well-mixed infinite
populations, defection is the evolutionarily stable strategy (ESS)
~\cite{Hofbauer98}, even though all individuals would be better
off if they cooperated. Thereby this creates the social dilemma,
because when everybody defects, the mean population payoff is
lower than that when everybody cooperates. Whereas in the SG, the
order of $P$ and $S$ is exchanged, such that $T>R>S>P$. Its
essential ingredient is that in contrast to the PDG, cooperation
has an advantage when rare, which implies that the replicator
dynamics of the SG converges to a mixed stable equilibrium where
both C and D strategies are present~\cite{Hofbauer98}. It is
important to note that in this state the population payoff is
smaller than it would be if everyone played C, hence the SG still
represents a social dilemma~\cite{Hauert04}. In addition, the SG
is of much applications and interests within biological context.
In order to solve these social dilemmas, a variety of suitable
extensions on these basic models has been
investigated~\cite{Nowak92,Doebeli05,Santos05,Szabo_review}. Most
importantly, it is found that cooperation can be promoted and
sustained in the network-structured population
substantially~\cite{Santos05,Santos2006a,Santos2006b}. Indeed, the
successful development of network science provides a convenient
framework for describing the dynamical interactions of games. The
evolution of cooperation on model networks with features such as
lattices~\cite{Szabo98,Szabo02,Szabo05,Vukov06},
small-world~\cite{Masuda03,Abra2001,Toma2006},
scale-free~\cite{Santos05}, and community structure~\cite{Chen06}
has been scrutinized. Moreover, the understanding of the effect of
network structure on the evolution of cooperation reaches to
consensus gradually: the heterogeneity of the network of contacts
plays a significant role in the emergence of cooperation. However,
the puzzle of cooperation is unanswered yet. What on earth
conditions the emergence of cooperation is still a challenging
problem~\cite{Pennisi05,Colman06}. Most noteworthy, Nowak
summarized five possible rules for the evolution of cooperation
corresponding to different situations (see Ref.~\cite{Nowak06} and
references therein). Nevertheless, to our best knowledge, these
results are mostly based upon simplified scenario and model. To
inspect the evolution of cooperation, further details and
characteristics of real world should be considered and integrated.

The World Wide Web (WWW) in its first decade was like a big online
library, where people mainly searched for information. Today,
owing to new social technologies, the web is undergoing a subtle
but profound shift, dubbed {\em Web 2.0}, to become more of a
social web, not unlike the WWW inventor Tim Berners-Lee's original
vision. The use of collaborative technologies such as blogs and
wikis also leads to change of the ways of people's thinking and
communicating. People, especially college students, take advantage
of online social network services for messaging, sharing
information, and keeping in touch with each other. This creates an
emerging online community which is largely being shaped by dynamic
interactions between users in real time. Therefore, these services
provide an extraordinary online laboratory to study dynamic
pattern of social interactions conveniently and effectively. In
what follows, the two aforementioned metaphors---PDG and SG in an
empirical social network will be examined.

In this paper, we present our observations into a Chinese social
networking site open to college students (the social networking
site Xiaonei, which began in late 2005 in select universities, but
grew quickly to encompass a very large number of universities.).
We empirically study the evolution of cooperation on such online
social network which retains the essential of real-world social
interactions. In particular, it is suggested that the evolution of
cooperation on empirical social networks is influenced by a variey
of network effects, including heterogeneity, small-world effect,
local highly-connected clusters, average connectivity, etc. In the
rest of this paper, first, we will analyze the structure of the
online social network, then investigate the two social dilemmas
(PDG and SG) on this social network by the method analogous to
replicator dynamics, observing the time evolution of cooperation.
After that, we discuss the simulation results and make
explanations. Finally, we draw the conclusion remarks.

\section{The structure of the online social network}
\begin{figure}
\centering
\includegraphics[width=10cm]{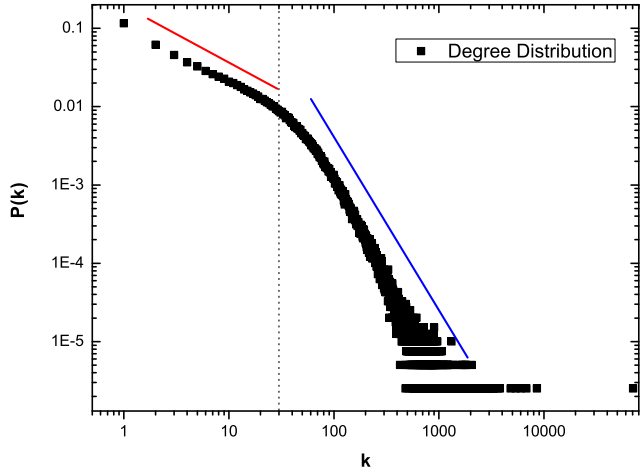}
\caption{The degree distribution $P(k)$ of the online social
network. The dot line indicates the critical degree $k_c=30$: for
$k<k_c$, $P(k)$ follows a power-law as $\sim k^{-0.72}$, while for
$k>k_c$, $P(k)$ obeys a power-law as $\sim k^{-2.12}$. The slopes
of the left and right straight lines are respectively $-0.72$ and
$-2.12$ for comparison with the degree distribution. \label{fig1}
}
\end{figure}

\begin{figure}
\centering
\includegraphics[width=10cm]{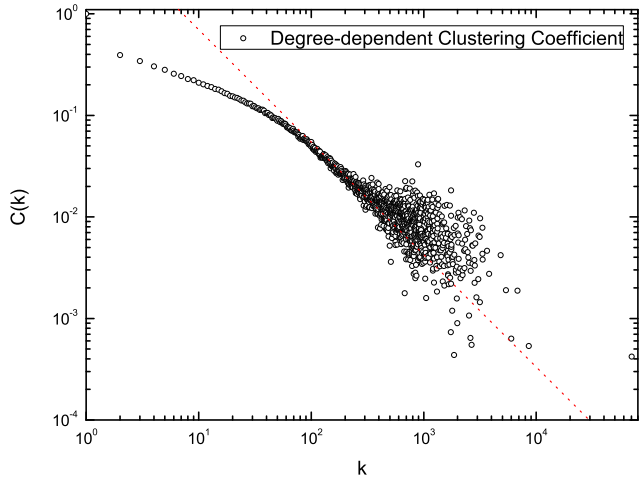}
\caption{The plot of degree-dependent clustering coefficient
$C(k)$ vs. degree $k$. A clear power-law is absent, but $C(k)$'s
dependence on $k$ is nontrivial. The dot line has slope
$-1$.\label{fig2}}
\end{figure}
The actual relational data was obtained from a Chinese social
networking site open to college students. We focus on the
connected component, which is composed of $396,836$ nodes and
$3,548,572$ bidirectional edges (we viewed this network as
undirected one). We perform statistical analysis of the structure
of this social network. The quantities such as degree
distribution, clustering coefficient, average shortest path
length, etc are calculated to capture the features of the online
social network. In Fig.~\ref{fig1}, we report the degree
distribution $P(k)$, which gives the probability that randomly
selected node has exactly $k$ edges. Clearly, we can see that
$P(k)$ follows two different scalings with $k$, depending on the
specified critical degree value $k_c$. $P(k)$ obeys a power-law
form $\sim k^{-\gamma_1}$ with $\gamma_1=0.72\pm 0.01$ when
$k<k_c=30$. Otherwise, $P(k)\sim k^{-\gamma_2}$, where
$\gamma_2=2.12\pm 0.02$ for $k>k_c$. The degree distribution above
the critical degree $k_c$ is consistent with past findings of
social networks with the degree exponent
$2<\gamma<3$~\cite{Albert_Barabasi_review_2002}. Whereas, for
small degree $k$ below $k_c$, the scaling exponent of $P(k)$ is
less than two. Considerable fraction of nodes have only low
connectivity (see Tab.~\ref{tab:degree}). About $68\%$ nodes'
degrees are not more than 30. The average degree $\langle k
\rangle$ is $17.9$.

\begin{table}
\caption{\label{tab:degree}Percentage of nodes with 1, 2, 3, 4,
and 5 degrees. Note that a large fraction of nodes have only small
degrees.}
\begin{tabular}{cccccc}
\hline
k= & 1 & 2 & 3 & 4 & 5 \\
\hline  & 11.6\% & 6.2\% & 4.6\% & 3.7\% & 3.3\%\\ \hline
\end{tabular}
\end{table}
The length of average shortest path $\langle l \rangle$ is
calculated, which is the mean of geodesic distance between any
pairs that have at least a path connecting them. In this case,
$\langle l \rangle = 3.72$. And the diameter $D$ of this social
networks which is defined as the maximum of the shortest path
length, is $12$. The clustering coefficient of node $i$ is defined
as $C_i=\frac{2E_i}{k_i(k_i-1)}$, that is the ratio between the
number $E_i$ of edges that actually exit between these $k_i$
neighbor nodes of node $i$ and the total number $k_i(k_i-1)/2$.
The clustering coefficient of the whole network is the average of
all individual $C_i$'s. We find the clustering coefficient
$C=0.27$, order of magnitude much higher than that of a
corresponding random graph of the same size
$C_{rand}=17.9/396836=4.51\times 10^{-5}$. Besides, the
degree-dependent local clustering coefficient $C(k)$ is averaging
$C_i$ over vertices of degree $k$. Fig.~\ref{fig2} plots the
distribution of $C(k)$ vs. $k$. For clarity, we add the dot line
with slope $-1$ in the log-log scale. However, it is hard to
declare a clear power law in our case. Nevertheless, the nonflat
clustering coefficient distributions shown in the figure suggests
that the dependency of $C(k)$ on $k$ is nontrivial, and thus
points to some degree of hierarchy in the networks. In many
networks, the average clustering coefficient $C(k)$ exhibits a
highly nontrivial behavior with a power-law decay as a function of
$k$~\cite{Ravasz03}, indicating that low-degree nodes generally
belong to well-interconnected communities (corresponding to high
clustering coefficient of the low-connectivity nodes), while
high-degree sites are linked to many nodes that may belong to
different groups (resulting in small clustering coefficient of the
large-degree nodes). This is generally the feature of a nontrivial
architecture in which small-degree vertices are well-clustered
around the hubs (high degree vertices), and organized in a
hierarchical manner into increasingly large groups. Thus, our
empirical social network has such fundamental characteristic of
hierarchy.

Another important element characterizing the local organization of
complex networks is the degree correlation of node $i$ and its
neighbor. Following Newman~\cite{degree-cor}, assortativity
coefficient $r$ is measured by the Pearson correlation coefficient
of the degrees at either ends of an edge, which can be written as
\begin{equation}
r=\frac{M^{-1}\sum_ij_ik_i-[M^{-1}\sum_i\frac{1}{2}(j_i+k_i)]^2}{M^{-1}\sum_i\frac{1}{2}(j_i^2+k_i^2)-[M^{-1}\sum_i\frac{1}{2}(j_i+k_i)]^2},
\end{equation}
where $j_i,k_i$ are the degrees of the vertices at the ends of the
$i$th edge, with $i=1,\cdots,M$ ($M$ is the total number of edges
in the observed graph or network). We calculate the degree
assortativity coefficient (or degree-degree correlation) $r$ of
the online social network. In our case, $r=-0.0036$, which means
the social network shows ``disassortative mixing'' on its degrees.
Networks with assortative mixing pattern are those in which nodes
with large degree tend to be connected to other nodes with many
connections and vice visa. Technical and biological networks are
in general disassortative, while social networks are often
assortatively mixed as demonstrated by the study on scientific
collaboration networks \cite{degree-cor}. Internet dating
community, a kind of social networks embedded in a technical one,
and peer to peer (P2P) social networks are similar to our case,
displaying a significant disassortative mixing pattern
\cite{Internet_dating,P2P}.

Herein, we have presented the structural analysis of our online
social network. The observed network has small-world property,
that is, high clustering coefficient and short average shortest
path length. Moreover, it is an inhomogeneous one, namely, the
tail of degree distribution obeys a power law. Additionally, it
has nontrivial hierarchical organizations---low-degree nodes
generally belong to well-interconnected clusters, while
high-degree vertices are linked to many nodes that may belong to
different groups. Besides, it exhibits disassortative mixing
pattern. In the successive section, we will investigate the
evolution of cooperation in the social network, revealing the
cooperation level is affected by the topological organizations of
the social network.

\section{Social dilemmas on the social network}
We consider the evolutionary PDG and SG on the sampled social
network (composed of $9,677$ nodes) which is a good representative
of the original large-scale one. Each vertex represents an
individual and the edges denote links between players in terms of
game dynamical interaction. The individuals are pure strategists,
following two simple strategies: cooperate (C) and defect (D). The
spatial distribution of strategies is described by a
two-dimensional unit vector for each player $x$, namely,
\begin{equation}
s=\left(\begin{array}{c}
  1 \\
  0 \\
\end{array}\right )\,\, \mbox{and}\,\,
\left(\begin{array}{c}
  0 \\
  1 \\
\end{array}\right ),
\end{equation}
for cooperators and defectors, respectively. Each individual plays
the PDG/SG with its immediate ``neighbors'' defined by their
who-meets-whom relationships and the incomes are accumulated. The
total income of the player at the site $x$ can be expressed as
\begin{equation}
P_x=\sum_{y\in \Omega_x}s_x^TMs_y,
\end{equation}
where the $\Omega_x$ denotes the neighboring sites of $x$, and the
sum runs over neighbor set $\Omega_x$ of the site $x$. Following
common practice~\cite{Nowak92,Hauert04,Szabo98,Nowak93}, the
payoff matrices have rescaled forms for PDG and SG respectively,
\begin{equation}
M=\left (\begin{array}{cc}
  1 \,& \,0 \\
  b \,& \,0 \\
\end{array}\right )\,\, \mbox{and}\,\,
\left (\begin{array}{cc}
  1 \,& \,1-r \\
  1+r \,& \,0 \\
\end{array}\right ),
\end{equation}
where $1<b<2$ and $0<r<1$.\\
In evolutionary games the players are allowed to adopt the
strategies of their neighbors after each round. Then, the
individual $x$ randomly selects a neighbor $y$ for possibly
updating its strategy. The site $x$ will adopt $y$'s strategy with
probability determined by the total payoff difference between
them~\cite{Szabo98}:
\begin{equation}
\label{transp} W_{s_x \leftarrow s_y} =
\frac{1}{1+\exp[(P_x-P_y)/T]},
\end{equation}
where the parameter $T$ characterizes the noise effects, including
fluctuations in payoffs, errors in decision, individual trials,
etc. $T=0$ denotes the complete rationality, in which the
individual always adopts the better strategy determinately.
Whereas $T\to\infty$ denotes the complete randomness of decision.
For finite value of $T$, it introduces bounded rationality to
individual's decision making.

In what follows, we present our investigations to the two social
dilemmas played by individuals occupying the vertices of the
sampled social network. The evolution of the frequency of
cooperators as a function of the parameter $b$ for PDG and $r$ for
SG is obtained. Besides, we also observe the time evolution of
cooperators under different values of $b$ ($r$). Initially, an
equal percentage of cooperators and defectors is randomly
distributed among the elements of the population. Here, we adopt
the synchronous updating rule. Each individual will adapt its
strategy according to Eq.~(\ref{transp}) after each round game.
Equilibrium frequencies of cooperators are obtained by averaging
over 1000 generations after a transient time of 20000 generations.
Each data point results from averaging over 100 runs. In the
following simulations, $T=0.02$ is kept invariant~\cite{note}.

\begin{figure}
\centering
\includegraphics[width=10cm]{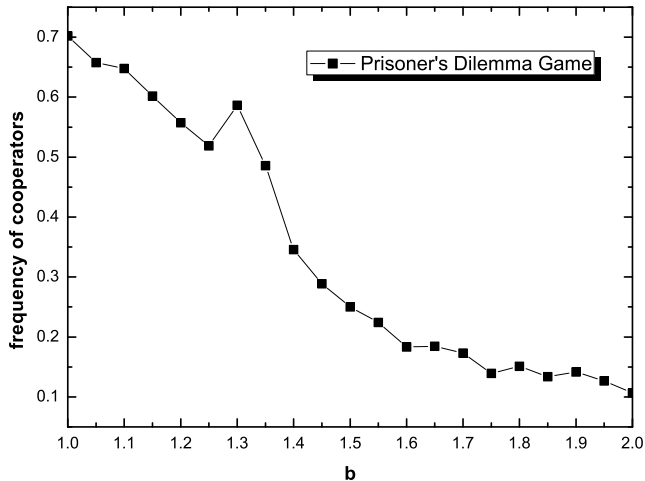}
\caption{Frequency of cooperators as a function of temptation to
defect $b$ in PDG.\label{fig3}}
\end{figure}

\begin{figure}
\centering
\includegraphics[width=10cm]{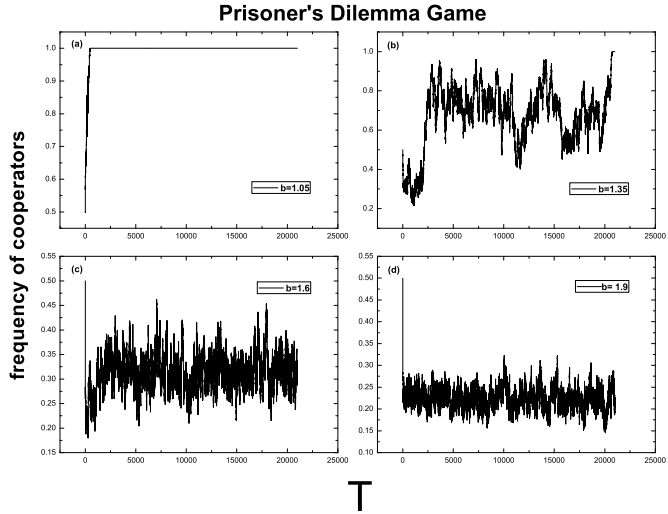}
\caption{The evolution of cooperation corresponding to different
values of $b$ in PDG. (a) $b=1.05$, (b) $b=1.35$, (c) $b=1.6$, (d)
$b=1.9$.\label{fig4}}
\end{figure}

In Fig.~\ref{fig3}, we report the frequency of cooperators $f_c$
as a function of temptation to defect $b$ in PDG. It is clear that
$f_c$ nontrivially decreases with increasing $b$. Note that the
cooperation level $f_c$ is not as remarkable as that in model
scale-free network [Barab\'{a}si-Albert (BA)], especially for
small $b$. Nonetheless, the cooperators can survive for the entire
range of $b$ ($1<b<2$), avoiding dying out, when confronted with
the intense invasion of advantaged defectors. Replicator dynamics
in well-mixed population points out defection is the only
evolutionarily stable strategy (ESS) in PDG. Hence cooperators
will be wiped out by natural selection in well-mixed populations.
In fact, interactions in real-world are heterogeneous, in the
sense that different individuals have different numbers of average
neighbors with whom they interact with, a feature associated with
a power-law dependence of the degree distribution. Previous study
on model BA scale-free network, which captures the real-world
heterogeneity, found that scale-free networks provide a unifying
framework for emergence of
cooperation~\cite{Santos05,Santos2006a,Santos2006b}. Here, our
empirical study also provides a convincing evidence that degree
heterogeneity is one of the factors promoting cooperation in
realistic social networks. It is shown the time evolution of
cooperation in PDG corresponding to different values of $b$ in
Fig.~\ref{fig4}. For small $b$ near one, starting from 50\%
cooperators, cooperators dominate the populations. When $b$
increases, the frequency of cooperation is diminished in a manner
that cooperation level drop rapidly down at first, after
generations and generations, the frequency of cooperation
struggles to recover to a higher level (see Panels (b) and (c) in
Fig.~\ref{fig4}). For large $b$ near two, although defectors
prevail in the networked populations, cooperators still can
survive in such tough environment (about 10\% on average).

\begin{figure}
\centering
\includegraphics[width=10cm]{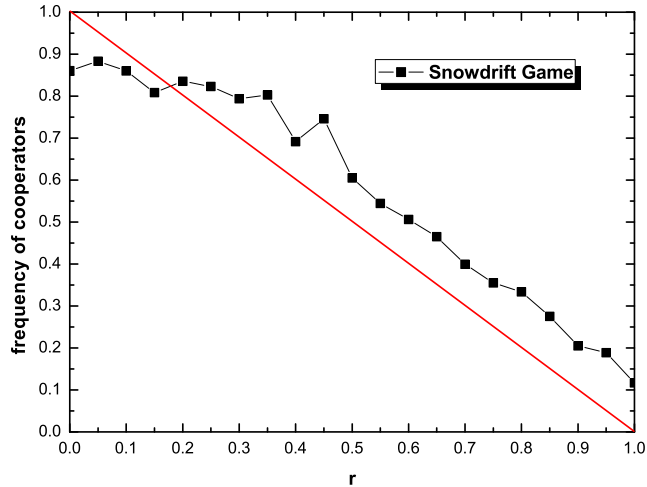}
\caption{Frequency of cooperators as a function of cost-to-benefit
ratio $r$ of mutual cooperation in SG.\label{fig5}}
\end{figure}

\begin{figure}
\centering
\includegraphics[width=10cm]{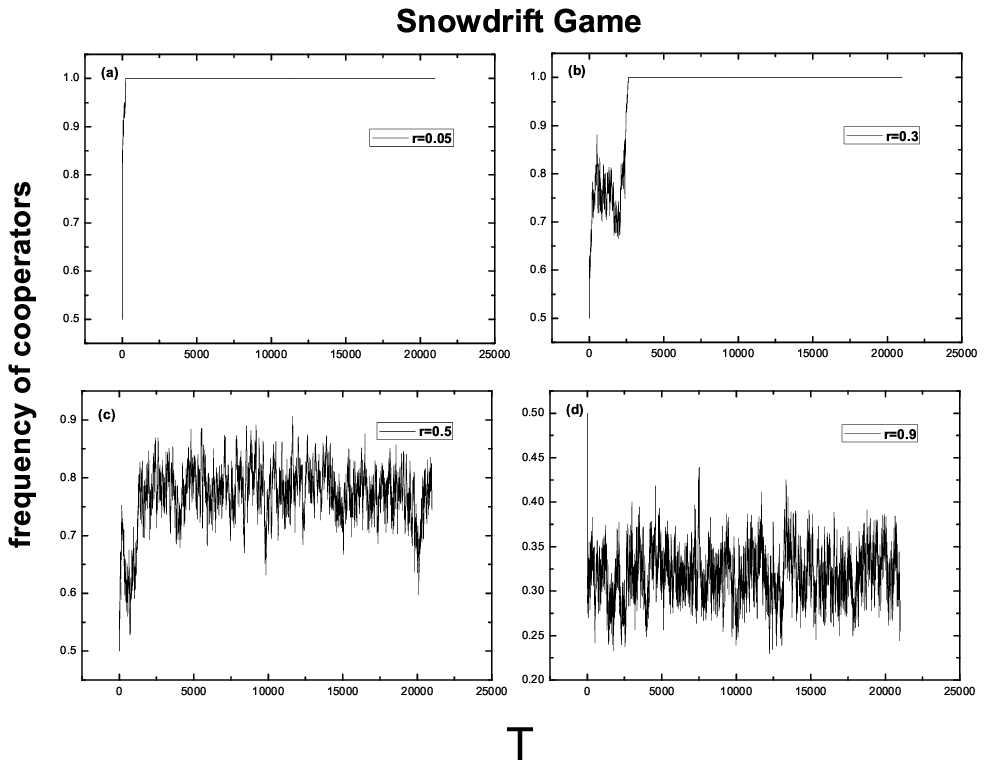}
\caption{The evolution of cooperation corresponding to different
values of $r$ in SG. (a) $r=0.05$, (b) $r=0.3$, (c) $r=0.5$, (d)
$r=0.9$.\label{fig6}}
\end{figure}

In parallel, the evolution of cooperation as a function of
cost-to-benefit ratio $r$ of mutual cooperation for SG is
presented in Fig.~\ref{fig5}.  In well-mixed scenario, replicator
dynamics of the SG converges to an equilibrium frequency for
cooperators given by $1-r$, corresponding to the red straight line
as shown in Fig.~\ref{fig5}. Except for small value $r$ near zero,
the cooperation level on our empirical social network is higher
than that in the well-mixed situation. Thus cooperation is
enhanced in our case. Generally speaking, because our social
network incorporates various features, including heterogeneity, of
which effects to cooperation are additive, the cooperation level
is not promoted as much as in model BA scale-free networks where
the cooperation level is mainly affected by heterogeneity. And
also, we investigated the time evolution of cooperation for SG, as
displayed in Fig.~\ref{fig6}. For small value of $r$, the
population frequently evolves into an absorbing state of full
cooperators when starting from an equal percentage of cooperators
and defectors [see Fig.~\ref{fig6}(a) and (b)]. As $r\to 1$, the
frequency of cooperators decreases from 50\% quickly, and
oscillates around the dynamic equilibrium state.

\begin{figure}
\centering
\includegraphics[width=10cm]{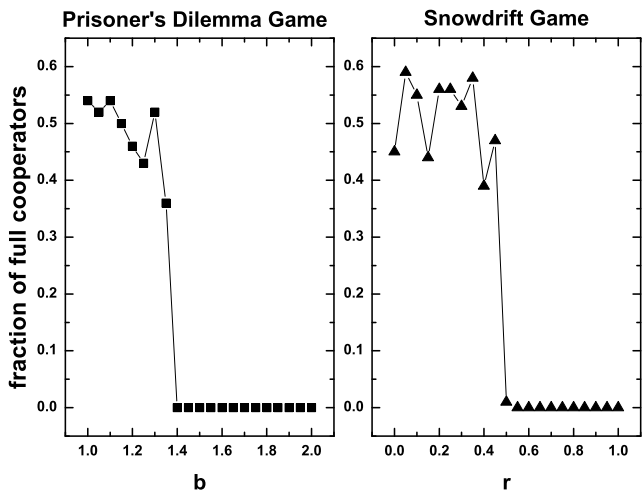}
\caption{Fraction of runs which ended with full cooperators vs.
$b$ ($r$) for PDG (SG). For each value of $b$ ($r$), we ran $100$
simulations, starting from $50\%$ cooperators.\label{fig7}}
\end{figure}

Moreover, fraction of runs which ended with full cooperators vs.
$b$ for PDG and $r$ for SG is shown in Fig.~\ref{fig7}. For each
value of $b$ and $r$, we ran 100 independent simulations, starting
from $50\%$ cooperators. We found that, for $b<1.4$ in PDG and
$r<0.5$ in SG, the networked population evolves into absorbing
state of full cooperators with a probability around 50\%. In this
situation, except for some individual runs ending up with full
cooperators, others ended up with low frequency of cooperation on
occasion as a result of the different initial distributions of
cooperators and defectors among the population. However, when
$b>1.4$ for PDG and $r>0.5$ for SG, most of the runs ended up with
massive defectors which resulted in the low frequency of
cooperation. Accordingly, even though the network structure
promotes cooperation, its positive influence to the evolution of
cooperation is to some extent suppressed by the increasing
parameters $b$ and $r$.

On the other hand, together with heterogeneity, other factors
including average connectivity~\cite{wwx}, small-world
effect~\cite{Masuda03}, degree-degree correlation~\cite{ZHRong},
randomness in topology~\cite{Renjie}, etc, play crucial roles in
the evolution of cooperation. Rather than investigations on model
networks where only one or few features of real-world
relationships are present, the evolution of cooperation on the
empirical social networks, which possess a variety of features in
topological organizations, should be understood from a
\emph{synthesis} view. In our case, besides the scale-free
feature, the online social network has small-world property,
hierarchical organizations and disassortative mixing pattern.
Concerning small-world property of the underlying network, the
short average distance promotes the propagation process of
cooperators. Furthermore, taking account for the hierarchical
organizations [due to $C(k)$'s dependence on $k$], i.e., local
well-clustered low-degree nodes, such common cluster structure
induces the clustering of cooperators, leading to the surviving
and enhancement of cooperation~\cite{Hauert04}. The mixing pattern
also influences the cooperation level substantially. It is thought
that the cooperation level is optimum in uncorrelated networks
(where the assortativity coefficient is zero)~\cite{ZHRong}. Our
sampled network is a disassortative one (assortativity coefficient
is $-0.0007$), thus the frequency of cooperators is diminished in
a way by such mixing pattern. Finally, as pointed out in
Ref.~\cite{wwx}, maximum cooperation level occurs at intermediate
average degree, in our case, the average connectivity of the
sampled network is about $10$. To a certain extent, this quantity
of average connectivity affects the evolution of cooperation.
Consequently, the evolution of cooperation on the empirical
network is simultaneously affected by these additive factors as
the underlying network possesses various characteristics of
real-world social interactions. Actually, the combined network
effects of these factors facilitate and maintain the cooperation
among selfish individuals. Our results may shed light on the
evolution of cooperation in social and economical systems.

\section{Concluding remarks}
In conclusion, we have studied two social dilemmas---PDG and SG in
an online social network. We demonstrated that the social network
has small-world property and is scale-free in degree distribution.
In addition, it was shown that the underlying network has
hierarchical organizations in which low-degree vertices are
well-connected in different communities, while large-degree nodes
are linked to many nodes that may belong to different groups. We
also found that the social network shows disassortative mixing
pattern. Then we investigated the evolution of cooperation on such
empirical social network, observing the time evolution of
frequency of cooperators for evolutionary PDG and SG respectively.
The underlying network structure leads to the enhancement and
maintenance of cooperation in unrelated and selfish individuals.
Besides, with not too large $b<1.4$ for PDG and $r<0.5$ for SG,
the networked population evolves into the absorbing state of full
cooperators at a probability around half one. Otherwise, most of
the runs ended up with massive defectors. Different from games on
model networks, understanding the evolution of cooperation on
empirical network should be conducted from a synthesis view
because real-world relational networks incorporate various
characteristics while model networks generally focus on some
specified features. Thus, we conclude that the evolution of
cooperation on the empirical network is jointly affected by
additive network effects, including average connectivity,
small-world property, degree heterogeneity (scale-free),
degree-degree correlation, hierarchical organizations, etc. Our
results may help understand the cooperative behaviors in human
societies.

\section*{Acknowledgement}
Delightful discussions with Wenxu Wang, Jing Wang, Zhuozheng Li,
and Zhoujin Ouyang are gratefully acknowledged. This work was
supported by NNSFC (60674050 and 60528007), National 973 Program
(2002CB312200), National 863 Program (2006AA04Z258) and 11-5
project (A2120061303).


\begin{thebibliography}{00}
\bibitem{note}\note{We have confirmed that our results are still qualitatively
valid under changing the value of $T$. However, the effect of
changing the value of $T$ is presented in Ref.~\cite{Renjie}.}
\end{thebibliography}
\end{document}